\documentclass[aps,prl,preprint]{revtex4}
\usepackage{graphicx}
\usepackage{dcolumn}
\usepackage{bm}
\usepackage{amssymb}
\usepackage{amsmath}
\usepackage{mathrsfs}
\usepackage{color}

\begin{document}

\bibliographystyle{unsrt}

\title{Boosted High Order Harmonics from Electron Density Singularity Formed at the Relativistic Laser Bow Wave}

\author{Jie Mu}
\email[]{Jie.Mu@eli-beams.eu}
\affiliation{Institute of Physics of the ASCR, ELI Beamlines Project, Na Slovance 2, 18221 Prague, Czech Republic}

\author{Timur Zh. Esirkepov}
\affiliation{Kansai Photon Science Institute, National Institutes for Quantum and Radiological Science and Technology, 8-1-7 Umemidai, Kizugawa, Kyoto 619-0215, Japan}

\author{Yanjun Gu}
\affiliation{Institute of Physics of the ASCR, ELI Beamlines Project, Na Slovance 2, 18221 Prague, Czech Republic}

\author{Tae Moon Jeong}
\affiliation{Institute of Physics of the ASCR, ELI Beamlines Project, Na Slovance 2, 18221 Prague, Czech Republic}

\author{Petr Valenta}
\affiliation{Institute of Physics of the ASCR, ELI Beamlines Project, Na Slovance 2, 18221 Prague, Czech Republic}
\affiliation{Czech Technical University in Prague, Faculty of Nuclear Sciences and Physical Engineering, Brehova 7, 11519 Prague, Czech Republic}

\author{Alexander S. Pirozhkov}
\affiliation{Kansai Photon Science Institute, National Institutes for Quantum and Radiological Science and Technology, 8-1-7 Umemidai, Kizugawa, Kyoto 619-0215, Japan}

\author{James K. Koga}
\affiliation{Kansai Photon Science Institute, National Institutes for Quantum and Radiological Science and Technology, 8-1-7 Umemidai, Kizugawa, Kyoto 619-0215, Japan}

\author{Masaki Kando}
\affiliation{Kansai Photon Science Institute, National Institutes for Quantum and Radiological Science and Technology, 8-1-7 Umemidai, Kizugawa, Kyoto 619-0215, Japan}

\author{Georg Korn}
\affiliation{Institute of Physics of the ASCR, ELI Beamlines Project, Na Slovance 2, 18221 Prague, Czech Republic}

\author{Sergei V. Bulanov}
\affiliation{Institute of Physics of the ASCR, ELI Beamlines Project, Na Slovance 2, 18221 Prague, Czech Republic}
\affiliation{Kansai Photon Science Institute, National Institutes for Quantum and Radiological Science and Technology,
8-1-7 Umemidai, Kizugawa, Kyoto 619-0215, Japan}

\date{\today}
\begin{abstract}
We demonstrate coherent hard electromagnetic 
radiation generation from reflection by the electron density singularity 
formed at the relativistic bow wave in laser plasma via particle-in-cell simulations. 
Wake and bow waves driven by an intense laser pulse form an electron 
density singularity at the laser pulse front where they join. 
A counter-propagating laser pulse is reflected at the electron density 
modulations moving with relativistic velocity. 
The reflected electromagnetic pulse is compressed and its 
frequency is upshifted. Its frequency spectrum contains relativistic 
harmonics of the driver pulse frequency generated at the bow wave 
front, all upshifted with the same factor as the fundamental mode of 
the incident light.

\end{abstract}

\maketitle

High brightness sources of electromagnetic radiation have attracted 
a great deal of attention due to the broad range of applications in biology, 
molecular imaging, material sciences and fundamental science 
research~\cite{suckewer1990soft, krausz2009attosecond, daido2002review, mourou2006optics}. 

One of the way towards developing ultra-short, intense 
electromagnetic pulse source is based on 
simultaneous laser frequency upshifting and
the pulse compression.  
These two phenomena were considered, in
particular, with the wave amplification reflected at the moving
relativistic electron slab in Ref. \cite{L52};
 the reflection at the moving ionization fronts studied in 
Refs. \cite{ION1, ION2, ION3, ION4}. 
A high repetition regime allowing one to produce frequency upshifted 
high intensity quasi-monochromatic electromagnetic radiation proposed 
in Ref.~\cite{bulanov2003light}, uses a laser produced breaking wake wave in 
underdense plasma as the flying mirrors to reflect, compress and focus the 
counterpropagating laser pulse 
(for details see review articles~\cite{bulanov2013relativistic, MK18} 
and references cited therein). This concept is based on the Einstein 
prediction~\cite{einstein1905electrodynamics} according to which, 
in the head-on wave-mirror collision, the reflected electromagnetic 
pulse is compressed with its frequency upshifted by a factor $4\gamma_M^2$.
Here,  $\gamma_M$ is the mirror Lorentz factor $\gamma_M=1/\sqrt{1-v_M^2/c^2}$ 
with $v_M$ and $c$ being the mirror velocity and speed of light in vacuum. 
 The flying mirror can be a dense plasma slab accelerated by a high 
contrast ultraintense laser pulse in the radiation pressure dominant 
regime~\cite{esirkepov2009boosted} or a laser accelerated 
thin electron layer~\cite{VVK07}. The underdense plasma with an 
up-ramp profile can lead to emission of electromagnetic pulses from 
laser wake fields under certain conditions~\cite{sheng2005emission}, 
and mitigate the premature wavebreaking 
due to thermal effects~\cite{mu2013robust}.
The oscillating mirrors formed as oscillating electron density 
modulations at the surface of an overdense plasma are used to 
generate high order harmonics~\cite{BNP94, NMN04, TG09}. 

As is known, a focused intense laser pulse propagating in underdense 
plasma excites wake waves~\cite{ESL09,  APMtV02, tajima1979laser}, 
in which electrons are pushed not only along the laser pulse propagation 
direction but also aside, creating a cavity void of electrons. The transverse 
motion of the electrons at the cavity walls leads to the transverse wake wave 
breaking~\cite{bulanov1997transverse}, resulting in the electron 
injection into the wake field accelerating phase. Due to the transverse 
electron motion at the laser pulse front the laser excites a bow 
wave~\cite{esirkepov2008bow} causing a large-scale transverse 
modulation of the electron density and electron singularities formed 
at the joint of the boundaries of  the cavity and bow wave. 
The electron singularity oscillations driven by the laser field generate 
high order harmonics. The harmonic frequency reaches the
``water window'' region as observed in the experiments on 
high power ultra-short pulse laser interaction with underdense 
plasmas and in computer simulations~\cite{pirozhkov2012soft, ASP17}.

\begin{figure}[h]
\centering
\includegraphics[width=0.8\textwidth]{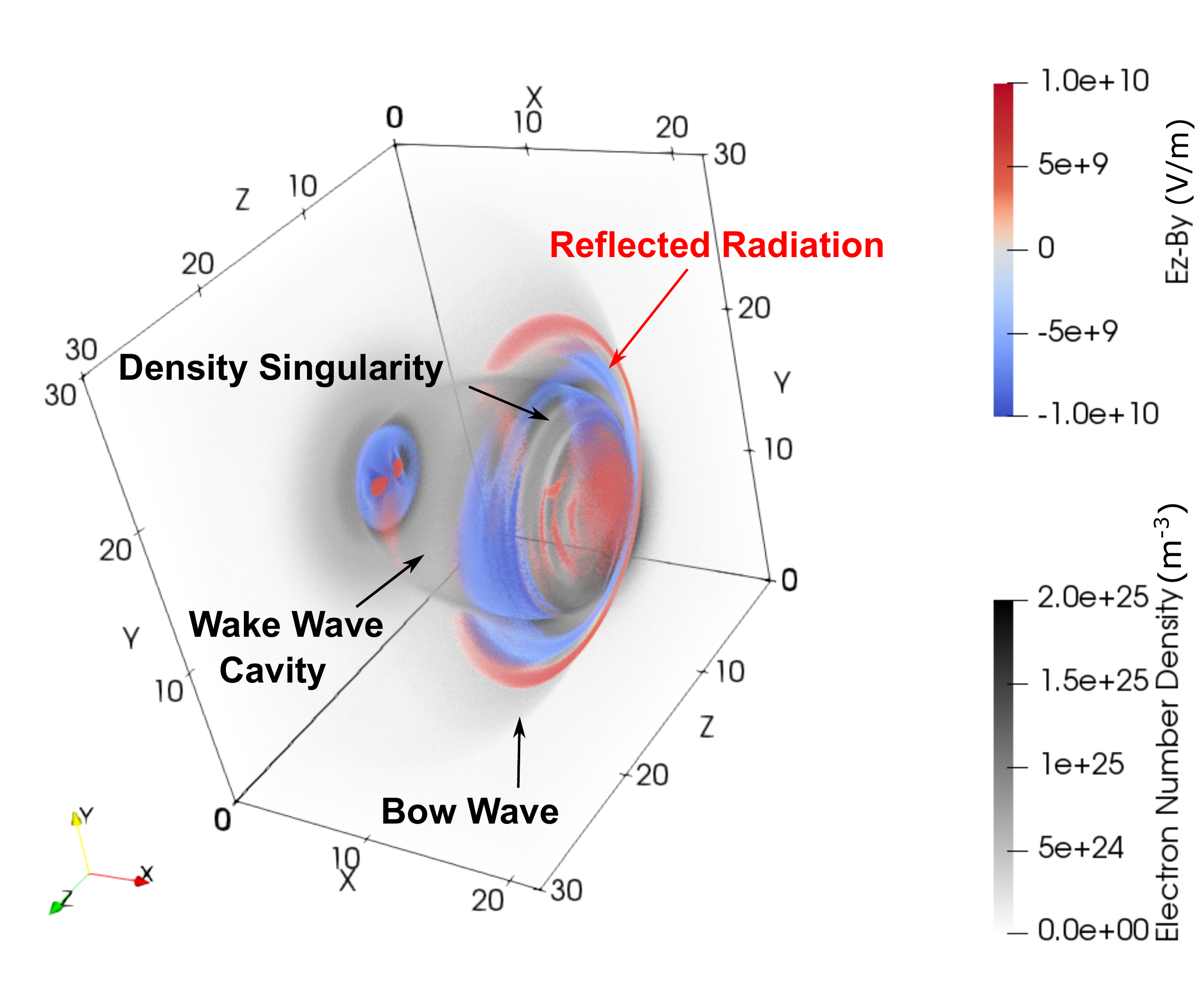}
\caption{(Color online) The wake wave cavity and bow wave with a singularity 
in the electron density distribution (in black and white) and the electromagnetic 
field reflected by the density spike and the cavity (in red and blue), at $t=24.3 T_d$.}
\label{Fig1} 
\end{figure}

In this letter, we propose a flying mirror scheme that uses the electron 
density singularity to reflect the counter-propagating laser pulse for 
laser frequency upshifting and for producing boosted high order harmonics. 
Within the framework of this scheme, first an intense driver laser pulse 
propagates through an underdense plasma to generate the wake and bow wave. 
At the region where the wake wave cavity wall joins the bow wave 
the electrons pile up to form a singularity in the electron density 
distribution moving with relativistic velocity. Second, a counter-propagating 
source laser pulse is reflected at the electron density singularity. 
The reflected electromagnetic pulse is compressed and its frequency 
is upshifted due the double Doppler effect. The frequency spectrum 
of reflected radiation contains relativistic harmonics generated at the 
bow wave front, all upshifted with the same factor as the fundamental 
mode of the incident light. We note that boosted high order harmonics 
have been seen in the spectrum of electromagnetic waves reflected by 
relativistic mirrors found with the computer simulations presented in 
Refs.~\cite{esirkepov2009boosted, koga2018relativisitcally}, 
when the relativistic mirrors were a high density plasma slab and a thin electron layer, 
respectively. In contrast to these cases, the configuration under 
consideration has the properties of a relativistic flying 
mirror~\cite{bulanov2003light}, of the oscillating relativistic mirror~\cite{BNP94}, 
and it inherits the properties of the laser driven oscillating electron spikes 
whose high efficiency in the high order harmonics generation is demonstrated
in Refs.~\cite{pirozhkov2012soft, ASP17}. Figure \ref{Fig1} shows 
the formation of the cavity and bow wave with a singularity in the 
electron density distribution, accompanied by the reflected 
electromagnetic field at $t=24.3 T_d$. Here $T_d= \lambda_d/c$ is 
the period of the driver laser and $\lambda_d$ is the wavelength of the driver laser.

\bigskip

\bigskip

To study two laser pulse interaction in the underdense plasma 
under the conditions, when the bow wave is formed, we carry out 
multi-dimensional particle-in-cell (PIC) simulation using the EPOCH 
code~\cite{arber2015contemporary}.

In the 3D simulations, the simulation box has the size of 
$22\lambda_d\times 30\lambda_d\times 30\lambda_d$. 
A spatial grid of $\Delta x/\lambda_d=1/30$, 
$\Delta y/\lambda_d=1/30$ and $\Delta z/\lambda_d=1/30$ is used to show the 
structure of the scheme and the reflected pulse. The fully ionized homogeneous 
density plasma slab is  located at $2\lambda_d \leq x \leq 22\lambda_d$, 
$0 \leq y \leq30\lambda_d$, and $0 \leq z \leq30\lambda_d$.
The electron density of the plasma is 
$n_e=1.14\times10^{19} ~{\rm cm}^{-3}\times (1 ~\mu m/\lambda_d)^2$, 
corresponding to $0.01n_c$. Here 
$n_c=m_e\omega^2/4\pi e^2=1.14\times10^{21}~ {\rm cm}^{-3}\times(1~\mu m/\lambda_d)^2$ 
is the critical plasma density, $e$ and $m_e$ are the charge and mass of electron, 
$\omega$ is the plasma frequency. The total number of the particles is 
$5.7\times10^8$. The ion response is neglected due to the large  ion 
to electron mass ratio and relatively low electron density.

We adopt the driver laser pulse with a normalized amplitude of 
$a_d=eE_d/m_e\omega_d c=6.62$, corresponding to the initial intensity 
equal to $I_d=6\times10^{19}\times(1~\mu m/\lambda_d)^2 ~{\rm W/cm^2}$. 
Here $E_d$ and $\omega_d$ are the electric field and frequency of the driver pulse, 
and $c$ is the speed of light in vacuum. The laser radiation is linearly 
polarized with the electric field directed along the $y$ axis. 
The full width at half maximum (FWHM) beam size is 
$5\lambda_d \times6.66\lambda_d\times6.66\lambda_d$. 
The driver laser pulse focus is at the left boundary of the simulation box.

\begin{figure}[h]
\centering
\includegraphics[width=0.8\textwidth]{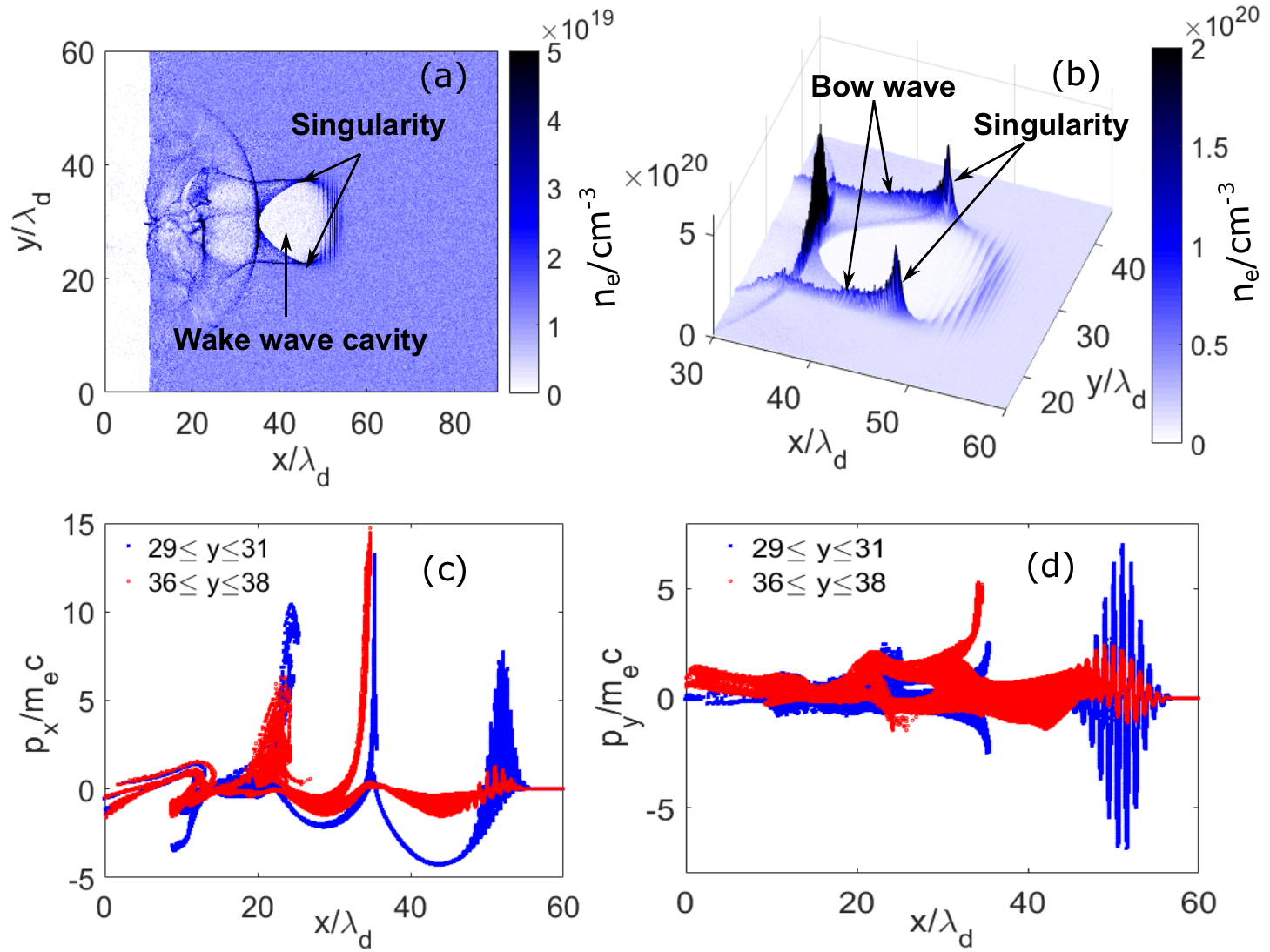}
\caption{
 (Color online)   Electron density in the $(x,y)$ plane in (a) 2 dimensional view 
and (b) 3 dimensional view at $t=54T_d$. (c) Longitudinal and (d) transverse 
electron  momentum $p_x$ and $p_y$ vs the $x$ coordinate, for either the 
particles in the central area $29\lambda_d \leq y \leq 31\lambda_d$ (in blue) 
or the ones in the upper singularity area $36\lambda_d \leq y \leq 38\lambda_d$ (in red).  }
\label{Fig2} 
\end{figure}

\begin{figure}[h]
\centering
\includegraphics[width=0.85\textwidth]{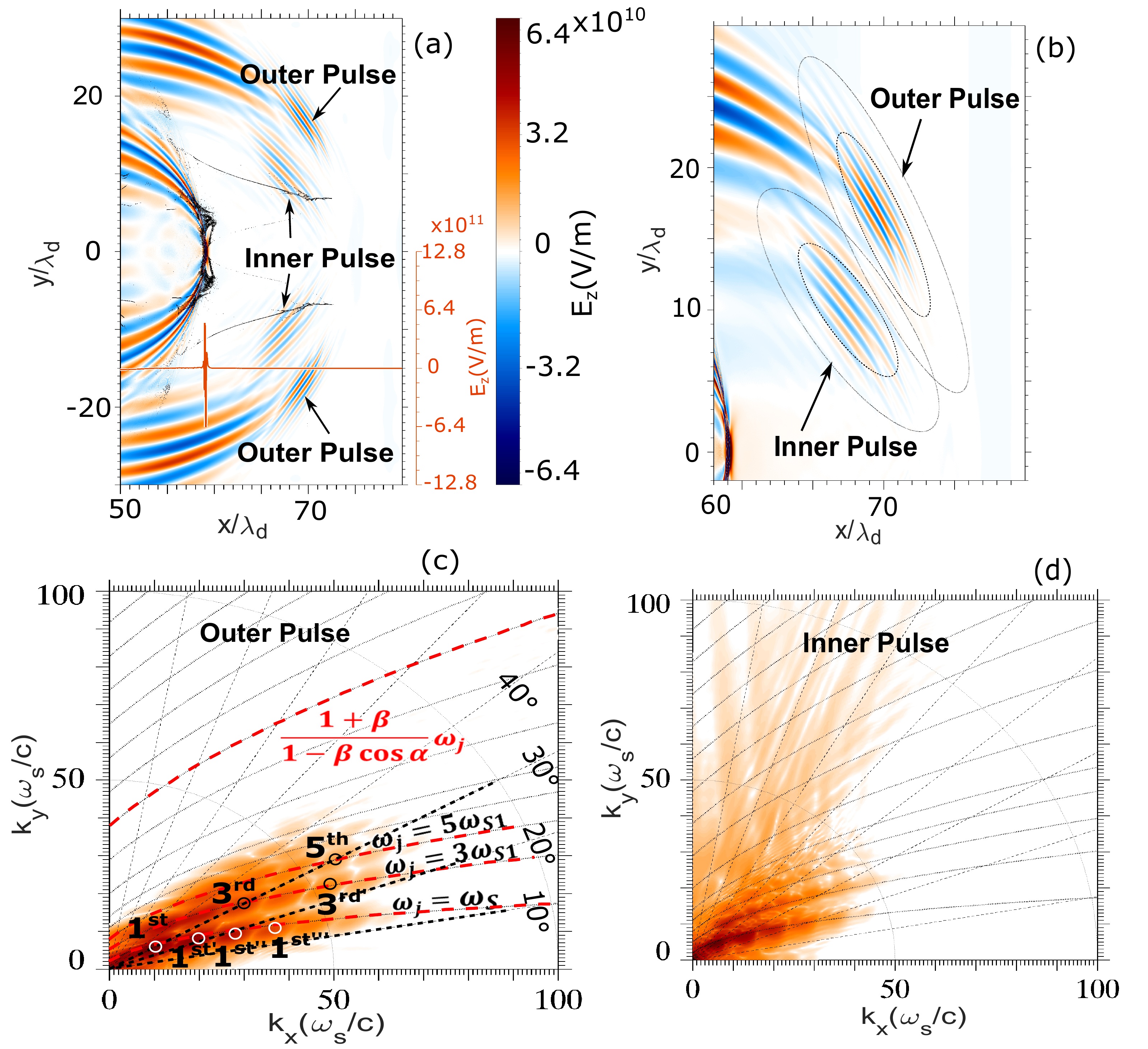}
\caption{ 
 (Color online) (a) The electric field $\rm E_z$ after high-pass filter, 
showing the frequency higher than the second harmonic of the source 
pulse $2\omega_s$, at $34.5T_d$. The black thin curves present 
the electron density isopleths at $0.03n_c$ and $0.04n_c$. The red 
thick curve presents $\rm E_z$ along the axis at $y=0$. 
(b) The outer and inner pulses of the reflected electromagnetic field, 
selected by a Gaussian spatial filter. (c) and (d) represent the frequency 
spectrum of the outer and inner pulses. Confocal ellipses marked with 
red dashed curve represent the frequency upshift dependence on the 
reflection angle $\alpha$. Straight lines marked with black dots represent 
the angle of the wave vector every 10 degrees. 
 }
\label{Fig3}
\end{figure}

The simulation results are shown in Figs. \ref{Fig1}, \ref{Fig2}, and \ref{Fig3}. 
Fig. \ref{Fig1} is the result of 3D simulation, and Figs. \ref{Fig2}, 
and \ref{Fig3} are the results of 2D simulations with similar parameters. 
In the 2D simulations, the simulation box has a larger size of 
$90\lambda_d\times 60\lambda_d$ to investigate the propagation of the 
reflected electromagnetic field. A substantially small-step spatial grid 
with  $\Delta x/\lambda_d=0.005$ and $\Delta y/\lambda_d=0.005$ is 
used to resolve the wavelength of the reflected pulse. The plasma slab 
is  located at $10\lambda_d \leq x \leq 90\lambda_d$, $0 \leq y \leq60\lambda_d$. 

Fig. \ref{Fig2} (a) shows the structure of the bow and wake waves, 
as well as the detailed view of the region where they join, 
i.e. of the region where the electron density singularity is formed at time $t=54T_d$. 
Note that the density singularity located in the $(x,y)$ plane is shown 
as two singularity points in the 2D simulation results 
illustrating the density in the $(x,y)$ plane. As shown in Fig. \ref{Fig2} (b), 
the electron density singularity has near-critical electron density. 
It is comparable to the electron density at the wake cavity bottom. 
Figs. \ref{Fig2} (c) and (d) display the longitudinal and transverse 
momentum $p_x$ and $p_y$ vs the coordinate $x$ for the 
particles in different regions. Compared with the particle density 
located in the central range of $29\lambda_d\leq y\leq31\lambda_d$, 
including injected fast electrons in the wake wave in blue, the particles 
around the density singularity area $36\lambda_d\leq y\leq38\lambda_d$ 
are shown in red to have lower momentum in the $x$ direction, 
but higher in the $y$ direction. The velocity of the density singularity 
is lower than the injected electrons, but still it is relativistic. 
Different from the $x$-axis symmetry of the $p_y$ distribution 
of the central particles, $p_y$ of the singularity particles are 
mostly above the $x$-axis, indicating that most of the 
singularity electrons move outside the wake cavity, 
and a small part of them moves downwards. 
A large number of the particles within the singularity
 have negative longitudinal momentum, but the particles 
localized near the driver laser front have large positive $p_x$. 

We note that the electron density singularity is observed to 
maintain stable structure and constant density for over more 
than 150 pulse cycles. The electron density in the singularity 
is approximately equal to the critical plasma density. 
The singularity moves with the velocity corresponding to 
substantially large relativistic factor $\gamma=1/\sqrt{n_c/n_e}=10$, 
where $n_e$ is the electron density of the plasma background. 
The velocity of the density singularity normalized to light 
speed is $\beta=v/c=\sqrt{1-1/\gamma^2}=0.995$.

Once the density singularity is generated, the source pulse irradiates 
it from the opposite direction to the driver laser pulse propagation. 
Another simulation is launched with a smaller grid of 
$\Delta x/\lambda_d=1/1024$ and $\Delta y/\lambda_d=1/256$ 
and a moving window. The source pulse is linearly polarized with 
the electric field $E_z$ directed along the $z$ axis. The driver and 
source pulses have different polarization for their radiation to be 
distinguished clearly from each other. The wavelength of the source 
pulse is longer than the driver pulse wavelength being equal to 
$\lambda_s=8\lambda_d$, here $\lambda_s$ is the wavelength 
of the source pulse. So that the reflected electromagnetic wave
 with the upshifted frequency can be more easily resolved for limited 
computing resources. The normalized amplitude of the source pulse 
is equal to $a_s=0.05$, corresponding to the intensity of 
$I_s=5.35\times10^{13}~{\rm W/cm^2}$. It is weak so as not to 
induce significant nonlinear response of the mirror electrons. 
The FWHM size of the source pulse is~$8\lambda_d\times33.3\lambda_d$, 
with  transverse size of the spot substantially large to guarantee the 
reflection of a significant amount of the photons at the density singularity. 
The source pulse is launched to encounter the singularity at $t=25T_d$.

The reflected electromagnetic field and its frequency spectrum are 
shown in Fig. \ref{Fig3}. Fig. \ref{Fig3} (a)  presents the reflected electric 
field $\rm E_z$ at $34.5T_d$, and the electron density isopleths at 
$0.03n_c$ and $0.04n_c$ at the same time. The source pulse is, first, 
partially reflected by the front part of the wake wave, and then immediately 
reflected by the density singularity, and after that the source pulse 
experiences the reflection from the bottom of the wake waves 
as in the normal flying mirror. The reflected radiation with the upshifted 
frequency shown in Fig. \ref{Fig3} (a) contains two parts reflected by the
two density singularity points in the $(x,y)$ plane. The interference 
of the two parts can also be seen. Each part of the reflected 
electromagnetic field contains two pulses, the outer pulse and the inner one. 
In this regime, the radiation is reflected from the density singularity 
which has multiple velocities and reflection angles according to the 
phase space shown in Figs. \ref{Fig2} (c) and (d). Thus, the outer 
and inner pulses reflected from different parts of the singularity 
are shown to have different frequency up-shift.

Fig. \ref{Fig3} (b) presents the upper part of the outer and inner pulses. 
The black thin curves show the spatial filter we use to select the 
reflected two pulses and perform the Fourier transformation. 
The energy of the outer electromagnetic pulse is estimated to be 
$2.1\times10^{-7} \rm~J$, which is 0.53\% of the source pulse. 
The number of photons is $6.4\times 10^{11}$, which is 
$4\times10^{-4}$ of the source pulse.

We first assume the density singularity as an inclined flat mirror 
moving with a dimensionless velocity of $\beta=v/c$.
Due to the double Doppler effect, the reflected pulse in 
$\rm E_z$ experiences a frequency upshift 

\begin{equation}
 \omega_r=\omega_s\frac{1+\beta\, {\rm cos}\,\phi}{1-\beta\, {\rm cos}\,\alpha} , 
\end{equation}

with the cosine of the the angle $\alpha$ between the reflected pulse and horizontal axis

\begin{equation}
 {\rm cos}\alpha=
 \frac{\left[2 \beta+(1+\beta^2){\cos}\, \phi\right]{\tan}^2\theta -2 {\tan}\,\theta \, {\sin}\,\phi-{\cos}\, \phi}
 {(1+\beta^2+2 \beta {\cos}\phi){\tan}^2\theta-2 \beta\, {\tan}\, \theta \, {\sin}\,\phi +1}. 
\end{equation}

Here $\omega_s$ and $\omega_r$ are the angular frequency of 
the source pulse and the frequency of the reflected pulse, 
$\phi$ is the angle between the propagation direction of 
the source pulse and the horizontal axis, $\theta$ is the 
angle between the normal to the mirror and the horizontal 
axis~\cite{landau1971classical}. In this regime, the source pulse 
propagates along the horizontal axis $x$. Using these relationships 
one can find the reflected pulse frequency
 
\begin{equation}
\omega_r=\omega_s\frac{1+\beta}{1-\beta\, {\cos}\,\alpha} .
\end{equation}

The Fourier transformed electric field $\rm E_z$ either for the outer 
and inner parts of the reflected pulse is presented in Figs. \ref{Fig3} (c) and (d). 
The peak of the reflected electromagnetic radiation propagates in a 
specific direction, which is determined by the singularity velocity 
and the tilt angle at the reflecting time. From the wave vector 
distribution in Fig. \ref{Fig3} (c), the angle between the reflected 
outer pulse and the horizontal axis, (i.e. reflection angle of the 
outer pulse) approximately equals $30^{\circ}$. So the frequency 
of the reflected radiation should be $\omega_r=14.4 \,\omega_s$ 
according to Eq. (3). From the simulation, we reach the maximum 
signal at the frequency  
\begin{equation}
\omega_r=c\sqrt{k_x^2+k_y^2}=\omega_{s1}\frac{1+\beta}{1-\beta\, {\cos}\,\alpha} \approx 11, 
\end{equation}
which corresponds to $\omega_{s1}\approx 0.76\, \omega_s$, here 
$\omega_{s1}$ is the frequency of the source pulse in plasma 
which is downshifted. This is caused by the depletion of the source 
pulse in plasma. The frequency downshift is more significant 
due to the low frequency of the source pulse. Nevertheless, 
the theoretical estimated frequency $\omega_r=14.4 \,\omega_s$ 
is also included in Fig. \ref{Fig3} (c) due to the wide distribution 
of the harmonic. 

The harmonics of the source pulse are boosted to higher frequency 
for both the outer and inner pulses. The different propagation directions 
of the harmonics represented by the dark areas located along the straight 
dotted grids in Fig. \ref{Fig3} (c), show the result of different reflection angles. 
The tilt angle of the mirror $\theta$ in this regime has a continuous 
range because the singularity is irregular with a curvature, instead of a flat mirror. 
Thus, the reflection angle also has a continuous range due to Eq. (2). 
The frequency upshift depends only on the mirror velocity and the reflection 
angle, as in Eq. (3). By substituting the reflection angle 
${\tan}\alpha = k_y/k_x$ and $\omega_r=\sqrt{k_x^2+k_y^2}$ into Eq. (3), 
we can obtain the canonical form of an ellipse in coordinates $(k_x,k_y)$ 
corresponding to the incident frequency $\omega_i$. The frequency of 
the reflected radiation will lie on the ellipses, as the black dashed curves 
show in Fig. \ref{Fig3} (c) and (d). Each ellipse stands for odd instances 
of $\omega_i$, corresponding to the harmonic orders.

The peaks of the frequency shown in white circles in  Fig. \ref{Fig3} (c) 
are lower than the analytical results of the ellipse corresponding to 
the frequency of the source pulse $\omega_s$, because in plasma 
the incident frequency is downshifted to 
$\omega_j=\omega_{s1}=0.76~\omega_s$ instead of $\omega_s$. 
The ellipses corresponding to odd instances of $\omega_{s1}$ 
are in good agreement with the peaks in the frequency 
spectrum of the simulation results. 

The reflected radiation has a different frequency spectrum from 
the harmonics generated by the density singularity itself. 
Both the self-produced harmonics and the boosted harmonics depend on 
the oscillating density singularity with periodic structure. 
The well-separated peaks of the frequency on the ellipses represent 
separated reflection angles, e.g. there are at least 6 different 
well-separated reflection angles for the first order harmonic 
$\omega_j=\omega_s$. This is due to the additive and destructive 
interference with respect to reflection angles, caused by the 
periodic curvature variation of the singularity.

Similar boosted high order harmonics are generated for the inner 
pulse, as shown in Fig. \ref{Fig3} (d). The main reflection angle 
of the inner pulse approximately equals $40^{\circ}$. There are at 
least 5 different well-separated reflection angles for the first order 
harmonic $\omega_j=\omega_s$. The frequency of the inner pulse 
is lower than that of the outer pulse in the figures, due to the
asymmetry of the transverse momentum $p_y$ on the $x$-axis. 

In conclusion, analyzing the properties of a two counterpropagating 
laser pulse interaction in underdense plasmas we proposed a novel 
scheme of the relativistic flying mirror for electromagnetic radiation 
frequency upshifting. The proposed scheme uses the laser pulse 
reflection at the electron density singularity moving with relativistic 
velocity. The singularity is formed in the region where the bow wave 
and the wake wave merge, producing the stable singularity mirror 
whose property is known from catastrophe theory~\cite{PoSt96}.

The source pulse is reflected by the electron density singularity 
as a flying mirror. The reflected electromagnetic wave has a 
frequency upshift due to the double Doppler effect. The frequency 
upshift depends on the mirror velocity and the mirror tilt angle 
with respect to its velocity direction.  High order harmonics are 
boosted to higher frequency with respect to various reflection 
angles, due to the periodic curvature variation of the singularity.

This scheme provides a promising way to produce ultra-bright 
radiation sources. It can also be used to investigate the dynamics 
in nonlinear physical processes in relativistic plasmas. The study 
on the electron density singularity geometry also contributes to 
the understanding of the electron motion in laser and underdense 
plasma interactions, especially the nature of the density singularity 
at the joining area of wake waves and bow waves. The electron 
density singularity as a relativistic electron mirror can be used 
to investigate black hole physics under laboratory 
conditions~\cite{chen2017accelerating}.

\begin{acknowledgments}

We appreciate discussions with  Mr. M. Matys. 
The work is supported by the project High Field Initiative 
(CZ.02.1.01/0.0/0.0/15\_003/0000449)
from the European Regional Development Fund, and the project 
``IT4Innovations National Supercomputing Center – LM2015070'' from 
The Ministry of Education, Youth and Sports in Czech Republic.  
JKK acknowledges support from JSPS KAKENHI Grant Number JP16K05639.

\end{acknowledgments}


\begin{thebibliography}{10}

\bibitem{suckewer1990soft} S. Suckewer and C. H. Skinner,
 {\em Science} {\bf 247},1553 (1990).

\bibitem{krausz2009attosecond} F. Krausz and M. Ivanov,
 {\em Rev. Mod. Phys.} {\bf 81}, 163 (2009).

\bibitem{daido2002review} H. Daido,
 {\em Reports on Progress in Physics} {\bf 65},1513 (2002).

\bibitem{mourou2006optics} G. A. Mourou, T. Tajima, and S. V. Bulanov,
 {\em Rev. Mod. Phys.} {\bf 78}, 309 (2006).
 
\bibitem{L52} K. Landecker, 
{\em Phys. Rev.} {\bf 86}, 852 (1952).

\bibitem{ION1} V. I. Semenova, 
{\em Sov. Radiophys. Quantum Electron.} {\bf 10}, 599 (1967).

\bibitem{ION2} W. B. Mori, 
{\em Phys. Rev. A} {\bf 44}, 5118 (1991).

\bibitem{ION3} R. L. Savage, Jr., C. Joshi, and W. B. Mori, 
{\em Phys. Rev. Lett.} {\bf 68}, 946 (1992).

\bibitem{ION4} A. Zhidkov, T. Esirkepov, T. Fujii, K. Nemoto, J. Koga, S. V. Bulanov, 
{\em Phys. Rev. Lett.} {\bf 103}, 215003 (2009).

\bibitem{bulanov2003light} S. V. Bulanov, T. Esirkepov, and T. Tajima, 
{\em Phys. Rev. Lett.} {\bf 91}, 085001 (2003). 

\bibitem{bulanov2013relativistic} S. V. Bulanov, T. Zh. Esirkepov, M. Kando, 
A. S. Pirozhkov, and N. N. Rosanov, {\em Physics Uspekhi} {\bf 56}, 429 (2013).
 
\bibitem {MK18} M. Kando, T. Zh. Esirkepov, J. K. Koga, A. S. Pirozhkov, and S. V. Bulanov, 
 {\em Quantum Beam Sci.} {\bf 2}, 9 (2018).
   
 \bibitem{einstein1905electrodynamics} A. Einstein,
 {\em Annalen der Physik} {\bf 17}, 891 (1905).

\bibitem{esirkepov2009boosted} T. Zh. Esirkepov, S. V. Bulanov, M. Kando, 
A. S. Pirozhkov, and A. G. Zhidkov, {\em Phys. Rev. Lett.} {\bf 103} 025002 (2009).
 
\bibitem{VVK07} V. V. Kulagin, V. A. Cherepenin, M. S. Hur, and H. Suk, 
{\em Phys. Plasmas} {\bf 14}, 113101 (2007).

 \bibitem{sheng2005emission} Z. M. Sheng, K. Mima, J. Zhang, and H. Sanuki,
 {\em Phys. Rev. Lett.} {\bf 94}, 095003 (2005).

\bibitem{mu2013robust}
J. Mu, F. Y. Li, M. Zeng, M. Chen, Z. M. Sheng, and J. Zhang,
{\em Appl. Phys. Lett.} {\bf 103}, 261114 (2013).

\bibitem{BNP94} S. V. Bulanov, N. M. Naumova, and F. Pegoraro, 
{\em  Phys. Plasmas} {\bf 1}, 745 (1994).

 \bibitem{NMN04} N. M. Naumova, J. A. Nees, I. V. Sokolov, B. Hou, and
G. A. Mourou, {\em Phys. Rev. Lett.} {\bf 92}, 063902 (2004).

\bibitem{TG09} U. Teubner and P. Gibbon, 
{\em Rev. Mod. Phys.} {\bf 81}, 445 (2009).

\bibitem{ESL09} E. Esarey, C. B. Schroeder, and W. P. Leemans, 
{\em Rev. Mod. Phys.} {\bf 81}, 1229 (2009).

\bibitem{APMtV02} A. Pukhov and J. Meyer-Ter-Vehn, 
{\em Appl. Phys. B} {\bf 74}, 355 (2002).

\bibitem{tajima1979laser}T. Tajima and J. M. Dawson, 
{\em  Phys. Rev. Lett.} {\bf 43}, 267 (1979).

\bibitem{bulanov1997transverse} S. V. Bulanov, F. Pegoraro, 
A. M. Pukhov, and A. S. Sakharov, {\em Phys. Rev. Lett.} {\bf 78}, 4205 (1997).

\bibitem{esirkepov2008bow} T. Zh. Esirkepov, Y. Kato, and S. V. Bulanov,
 {\em Phys. Rev. Lett.} {\bf 101}, 265001 (2008).

\bibitem{pirozhkov2012soft}
A. S. Pirozhkov, M. Kando, T. Zh. Esirkepov, P. Gallegos, H. Ahmed, E. N. Ragozin, A. Y.
  Faenov, T. A. Pikuz, T. Kawachi, A. Sagisaka {\em et al.}, {\em Phys. Rev. Lett.} 
{\bf 108}, 135004 (2012).
 
 \bibitem{ASP17} A. S. Pirozhkov, T. Zh. Esirkepov, T. A. Pikuz, A. Ya. Faenov, 
K. Ogura, Y. Hayashi, H. Kotaki, E. N. Ragozin, D. Neely, H. Kiriyama {\em et al.}, 
 {\em Sci. Rep.} {\bf 7}, 17968 (2017).

\bibitem{koga2018relativisitcally} J. K. Koga, S. V. Bulanov, T. Zh. Esirkepov, 
M. Kando, S. S.   Bulanov, and A. S.  Pirozhkov,
 {\em Plasma Phys.  Contr. Fus.} {\bf 60}, 074007 (2018).

\bibitem{arber2015contemporary}
T. D. Arber, K. Bennett, C. S. Brady, A. L. Douglas, M. G. Ramsay, N. J. Sircombe,
  P. Gillies, R. G. Evans, H. Schmitz, A. R. Bell {\em et al.},
 {\em Plasma Phys.  Contr. Fus.} {\bf 57},113001 (2015).

\bibitem{landau1971classical} L. D. Landau and E. M. Lifshitz, 
{\it The classical theory of fields} (Pergamon Press, Oxford, 1971).

\bibitem{PoSt96} T. Poston and I. Stewart, 
{\it Catastrophe Theory and Its Applications} (Dover, New York, 1996).

\bibitem{chen2017accelerating} P. Chen and G. Mourou, 
 {\em  Phys. Rev. Lett.} {\bf 118}, 045001 (2017).


\end{thebibliography}
\end{document}